\documentclass[runningheads]{llncs}
\usepackage{graphicx}
\usepackage{mathtools}
\usepackage{booktabs}
\usepackage{dirtytalk}
\newcommand{\repeatthanks}{\textsuperscript{\thefootnote}}

\begin{document}
%
\title{AutoSeg - Steering the Inductive Biases for Automatic Pathology Segmentation}
\titlerunning{AutoSeg}
%
\author{Felix Meissen\inst{1,2} \and
Georgios Kaissis\inst{1,2,3}\thanks{Equal contribution} \and
Daniel Rueckert\inst{1,2,3}\repeatthanks}
\authorrunning{F. Meissen et al.}
%
\institute{Technical University of Munich (TUM), Munich, Germany \and
Klinikum Rechts der Isar, Munich, Germany \and
Imperial College London, UK \newline
\email{felix.meissen@tum.de}}
%
\maketitle              
\begin{abstract}
In medical imaging, un-, semi-, or self-supervised pathology detection is often approached with anomaly- or out-of-distribution detection methods, whose inductive biases are not intentionally directed towards detecting pathologies, and are therefore sub-optimal for this task.
To tackle this problem, we propose AutoSeg, an engine that can generate diverse artificial anomalies that resemble the properties of real-world pathologies.
Our method can accurately segment unseen artificial anomalies and outperforms existing methods for pathology detection on a challenging real-world dataset of Chest X-ray images.
We experimentally evaluate our method on the Medical Out-of-Distribution Analysis Challenge 2021.
\footnote{Code available under: \url{https://github.com/FeliMe/autoseg}}

\end{abstract}
\section{Introduction}

Anomalies are samples that deviate from a predefined norm.
In medical images, these can manifest in various ways. Inaccuracies in image acquisition -- like motion artifacts in MRI -- can be considered as anomalies, as well as pathologies like tumors, natural intra-patient variations, or images from other modalities, such as natural images.
In most applications however, we are interested in detecting pathologies.
Therefore, the problem is ill-defined, and detection methods that follow this broad definition are likely to struggle with this difficult task.
Moreover, all anomaly detection models have some form of inductive bias, making certain types of anomalies harder to detect than others. Especially reconstruction-based anomaly detection methods have a strong inductive bias because of the scoring function that is based on relative intensity differences between the original image and the reconstruction.
However, the inductive biases of many methods are not steered towards that problem.
Recent works have shown that machine learning models can successfully be trained on synthetic data \cite{autoflow,fakeit}. Apart from performance improvements, this allows for more control over the models' detection characteristics.
We therefore, present a new approach that creates a useful inductive bias to better detect pathologies in medical images.

Our contributions are the following:
\begin{itemize}
    \item We propose AutoSeg, a novel strategy for generating artificial anomalies that represent characteristics of real-world anomalies better than existing methods.
    \item We evaluate our approach on multiple modalities including a challenging and unsolved real-world dataset.
    \item We achieve state of the art detection performance on artificial and real-world anomalies.
\end{itemize}

\section{Related Work}

We sort related work in medical anomaly detection into three categories:

Reconstruction-based methods train a generative model -- such as an Autoencoder (AE) or a Generative Adversarial Network (GAN) -- on images from healthy subjects only. This way, the model learns the underlying distribution and will fail to reconstruct regions in images that are anomalous, and were thus not observed during training.
Prominent representatives of these methods are \cite{BrainLesionAE,f-AnoGAN}. In \cite{BrainLesionAE}, the authors use a convolutional Autoencoder as their generative model.
In \cite{f-AnoGAN}, Schlegl \textit{et al.} train a GAN to represent the \say{healthy} distribution and during test-time use restoration to find an image that is both close to the input image and the learned manifold.
Recently, Baur \textit{et al.} \cite{ComparativeStudy} compared all reconstruction-based methods in a large study. We refer the reader to their work for a more thorough overview thereof.
This family of methods, however, has a very strong inductive bias, and it was recently shown by Meissen \textit{et al.} \cite{meissen} that for brain MRI, it can be outperformed via simple thresholding.

The second category contains methods that attempt to directly evaluate the likelihood of a sample being from the \say{healthy} distribution. They also use generative models trained on images from healthy patients only to compute the likelihood.
Pinaya \textit{et al.} \cite{transformer} train an ensemble of autoregressive transformers on the latent space of a pretrained fully-convolutional Vector Quantised-Variational Autoencoder (VQ-VAE) to estimate the likelihood of every spatial feature in this latent space.
Zimmerer \textit{et al.} \cite{BrainVAE} use the gradient of the evidence lower bound (ELBO) of a trained Variational Autoencoder (VAE) to detect anomalies in brain MRI. The ELBO is a lower bound on the actual log-likelihood of a sample $x$: $\operatorname{ELBO}(x) \leq \log p(x)$.

Recently, training on synthetic data became a popular approach for machine learning in regimes where annotated data is hard to acquire. It has been successfully applied to optical flow estimation \cite{autoflow} and face-related computer vision \cite{fakeit}.
In medical imaging, Tan \textit{et al.} \cite{fpi} trained a self-supervised segmentation model to detect artificial anomalies on brain MRI and abdomen CT.
The anomalies are created by selecting two images from different patients and interpolating between them at randomly sampled rectangular patches with random interpolation factors.

\section{Method}

\begin{figure}[htpb]
    \centering
    \includegraphics[width=1.0\textwidth]{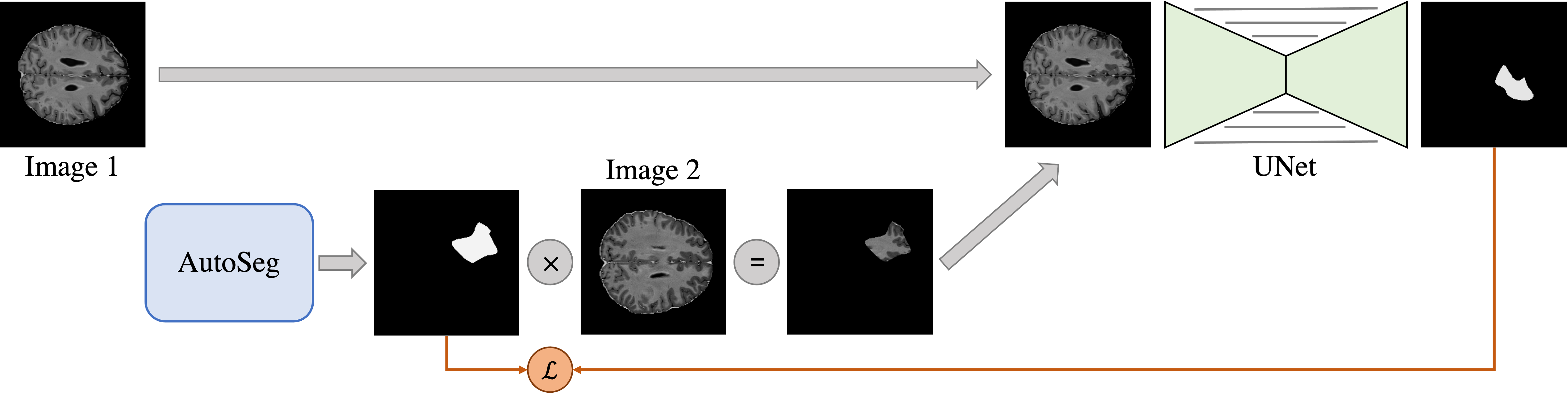}
    \caption{Overview of our method. An anomaly mask is generated via AutoSeg. The anomalous texture is taken from a different image like in \cite{fpi}. A UNet predicts the anomaly mask and strength of interpolation.}
    \label{fig:overview}
\end{figure}

\subsubsection*{Baseline}

We build our method upon the work of Tan \textit{et al.} \cite{fpi}.
To create artificial anomalies, we choose two images at random, sample a rectangular mask at a random location inside the two, uniformly sample an interpolation factor in the interval $[0.05, 0.95]$, and interpolate the images at the sampled mask with the interpolation factor. A segmentation model is then trained to segment the anomalous region, as well as to predict the interpolation factor.
Unlike the original work, we use a UNet \cite{pytorchunet} as our segmentation model.

\subsubsection*{AutoSeg}

To increase the realism of the artificial anomalies, we propose AutoSeg, an engine to generate anomalies with diverse characteristics.
AutoSeg is able to generate a specified number of anomalies in the shape of random polygons with a controllable number of vertices. This way, single large anomalies can be created -- as it is common for tumors -- or multiple smaller ones like in multiple sclerosis (MS) lesions.
The artificial anomaly generation process consists of two parts: Creating the anomaly mask, and choosing the anomalous texture. We use our proposed AutoSeg for the former and adhere to the patch interpolation strategy of the baseline for the latter.
Figure \ref{fig:overview} shows an overview of our method.

\subsubsection*{Volumetric Data}

Medical data is often three-dimensional. Despite that, most existing methods for anomaly detection are processing single slices independently, discarding all spatial information along one axis \cite{BrainLesionAE,f-AnoGAN,BrainVAE,transformer,fpi,ComparativeStudy}.
Using 3D convolutions is a potential solution to this problem but comes with significant computational costs.
This forces the user to either downsample the data or to use a patch-based approach.
Both alternatives limit the spatial resolution along all three axes.
Our solution mimics how radiologists look at volumetric data and is similar to work of Perslev \textit{et al.} \cite{acs_training}.
Instead of providing only one viewing direction, we train our models with samples from all three viewing directions (axial, coronal, and sagittal) equally as in.
During test-time, we perform inference on all three viewing directions and fuse the results by taking the average across all three predictions.
This allows us to use the same architecture for 2D and 3D data and requires only minimal changes to the training scheme.
We also feed $k=3$ adjacent slices at once to the segmentation network as channel dimensions to further increase the spatial information.

\section{Evaluation} \label{sec:evaluation}

\subsubsection*{Datasets}

We evaluate our method on multiple publicly available data sets.
First, we apply our method to the Medical Out-of-Distribution Analysis Challenge 2021 (MOOD) \cite{mood}.
The data contains 800 brain T2 MRI scans of healthy young adults from the Human Connectome Project \cite{hcp} with $256 \times 256 \times 256$ pixels per scan, and 550 abdominal CT scans of patients over 50 years of age \cite{moodct} with $512 \times 512 \times 512$ pixels per scan.
We split the two data sets into $60\%$ training and $40\%$ test data. From the training set, we use $5\%$ to evaluate performance during training.
Since no test data is publicly available for the MOOD challenge, we create our own artificial anomalies for evaluation on $75\%$ of the held-out test set.
For every anomalous sample $A$, we chose a sphere $h$ in the scan at random and add one of eight anomaly types to the pixels in the sphere.
We use the six anomaly types of Tan \textit{et al.} \cite{fpi} and add two more, called local blur and slice shuffle.
For the local blur, we create a blurred version $A'$ of the input scan via Gaussian filtering and replace the original sample $A$ with $A'$ at the anomalous region $h$.
For the slice shuffle, we choose a random axis and replace each slice along this axis in the patch with another slice inside that patch.


To test the suitability of our method for clinical applications, evaluation on a non-trivial real-world dataset is necessary.
Therefore, we further evaluate our method on the publicly available ChestX-ray14 dataset \cite{crx14}, containing $112,120$ frontal-view X-ray images of $30,805$ patients with 14 disease labels.
The dataset also includes the disease bounding boxes for 984 images.
We only consider $43,322$ images of patients over 18 and with posteroanterior view from which we use $75\%$ for training and the rest for evaluation.
While the original image resolution is $1024 \times 2014$, we bilinearly downsample all images to $256 \times 256$ -- maintaining the aspect ratio -- to be in line with the brain MR images.
Because of the obvious domain shift, we processed the male and female patients separately.
This dataset is challenging for pathology detection, because of the large intra-patient and intra-image variance in the patients' position, their anatomy, and the presence of external objects such as pacemakers.

\subsubsection*{Experimental Setup}

We implement our method in the PyTorch \cite{pytorch} framework. For the UNet, we choose the implementation of Buda \textit{et al.} \cite{pytorchunet}.
We train our method for 5 epochs using the AdamW optimizer \cite{adamw} with the default parameters for the UNet, a batch size of 8, and a learning rate of $0.0001$.
For the ChestX-ray14 dataset, we choose a larger batch size of $64$, a smaller model with width = 16, and train for 10 epochs.
We compare our method against a reconstruction-based VAE and report the average precision for every experiment.
For reference, we also include the performance a randomly guessing model would achieve.
AutoSeg is tuned to generate single anomalies with $10$ vertices, cubic spline interpolation between the vertices, and sizes uniformly sampled between $10\%$ and $50\%$, $20\%$ and $60\%$, and $5\%$ and $70\%$ of the image size for brain MRI, abdominal CT, and Chest X-ray respectively.
We deliberately performed only minimal hyperparameter tuning -- varying only the model width and learning rate -- to emphasize the contribution of our method over optimization strategy improvements.
\section{Results} \label{sec:results}

\begin{table}[htpb]
\caption{Average precision of our and comparing methods on unseen artificial anomalies of the held-out test set.}
\label{tab:mood_results}
\centering
\bgroup
\begin{tabular}{lcccc}
\toprule
               & \multicolumn{2}{c}{\textbf{Brain MRI}} & \multicolumn{2}{c}{\textbf{Abdomen CT}}   \\
\midrule
\textbf{Method}& \textbf{Sample}    & \textbf{Pixel}    & \textbf{Sample}     & \textbf{Pixel}      \\
\midrule
Random          & 0.750              & 0.003             & 0.750               & 0.004              \\
VAE             & 0.740              & 0.016             & 0.749               & 0.011              \\
UNet (Baseline) & 0.956              & 0.427             & -                   & -                  \\
+AutoSeg        & 0.999              & 0.954             & -                   & -                  \\
+2.5D Training  & \textbf{1.000}     & \textbf{0.974}    & \textbf{0.988}      & \textbf{0.953}     \\
\bottomrule
\end{tabular}
\egroup
\end{table}

Here we show the results of the evaluation described in Section \ref{sec:evaluation}. 
In Table \ref{tab:mood_results} the segmentation and detection performance of our models on the artificial anomalies from the held-out test set is presented.
Using our proposed AutoSeg yields the strongest performance improvement, leading to almost perfect segmentation. The worst performance in this experiment is achieved by the VAE, not substantially exceeding random guessing.

\begin{table}[htpb]
\caption{Average precision for pixel-wise evaluation on MOOD brain MRI.}
\label{tab:mood_brain_pixel}
\centering
\bgroup
\begin{tabular}{ccccc}
\toprule
\multicolumn{1}{l}{} & \textbf{VAE} & \textbf{UNet(Baseline)} & \textbf{+AutoSeg} & \textbf{+2.5D Training} \\
\midrule
Local Blur           & 0.012        & 0.915             & 0.997              & \textbf{0.999}           \\
Slice Shuffle        & 0.008        & 0.134             & 0.930              & \textbf{0.990}           \\
Noise Addition       & 0.015        & 0.124             & 0.908              & \textbf{0.936}           \\
Reflection           & 0.018        & 0.451             & 0.986              & \textbf{0.989}           \\
Sink Deformation     & 0.017        & 0.045             & 0.922              & \textbf{0.958}           \\
Source Deformation   & 0.016        & 0.550             & 0.944              & \textbf{0.968}           \\
Uniform Addition     & 0.022        & 0.699             & \textbf{0.987}     & 0.986                    \\
Uniform Shift        & 0.021        & 0.603             & 0.961              & \textbf{0.967}           \\
\textbf{Total}       & 0.016        & 0.427             & 0.954              & \textbf{0.974}           \\
\bottomrule
\end{tabular}
\egroup
\end{table}

In Table \ref{tab:mood_brain_pixel}, we present a detailed evaluation of the different artificial anomaly types. While performance on the UNet baseline is greatly different between different anomaly types, the models trained with AutoSeg show comparable average precision scores among all types.
Slice shuffle anomaly benefits the most from using 2.5D information during training. This is expected behavior, as it is the only anisotropic anomaly and is easier to detect from some viewing directions than from others.
Although all other anomalies are isotropic, 2.5D training also helps in these cases.
Additionally, we evaluate our final model trained with AutoSeg and 2.5D information on the MOOD 2021 challenge \cite{mood}. Here, the model achieves 6th place in the sample-level, and 4th in the pixel-level evaluation.

Table \ref{tab:cxr14_results} only shows sample-wise results, as there are no anomaly segmentations included in the ChestX-ray14 dataset. Here, the performance of our model trained with AutoSeg is substantially lower compared to the artificial anomalies but still outperforms all competing methods.

\begin{table}[htpb]
\caption{Sample-wise average precision of our and comparing methods on posteroanterior images of patients over 18 from the ChestX-ray14 dataset, split by gender.}
\label{tab:cxr14_results}
\centering
\bgroup
\begin{tabular}{lcc}
\toprule
                & \multicolumn{1}{c}{\textbf{Male}} & \multicolumn{1}{c}{\textbf{Female}}   \\
\midrule
Random          & 0.561 & 0.579 \\
VAE             & 0.607 & 0.624 \\
UNet(Baseline) & 0.539 & 0.576 \\
+AutoSeg        & \textbf{0.643} & \textbf{0.647} \\
\bottomrule
\end{tabular}
\egroup
\end{table}

\begin{figure}[htpb]
    \centering
    \includegraphics[width=0.8\textwidth]{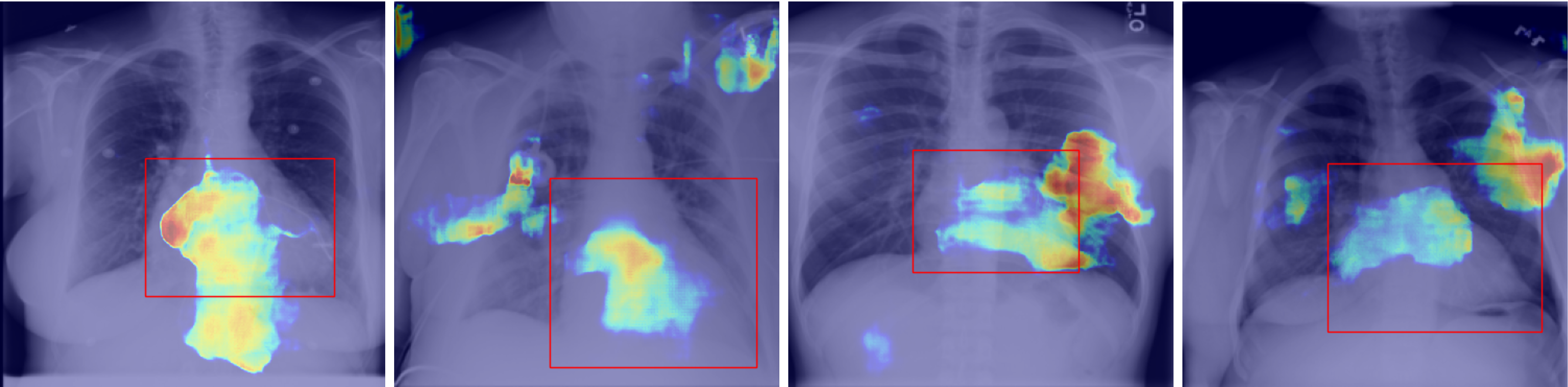}
    \caption{Selected examples of patients with cardiomegaly overlaid with the predicted anomaly map and the ground truth bounding box.}
    \label{fig:cxr14_cardiomegaly}
\end{figure}

\begin{figure}[htpb]
    \centering
    \includegraphics[width=1.0\textwidth]{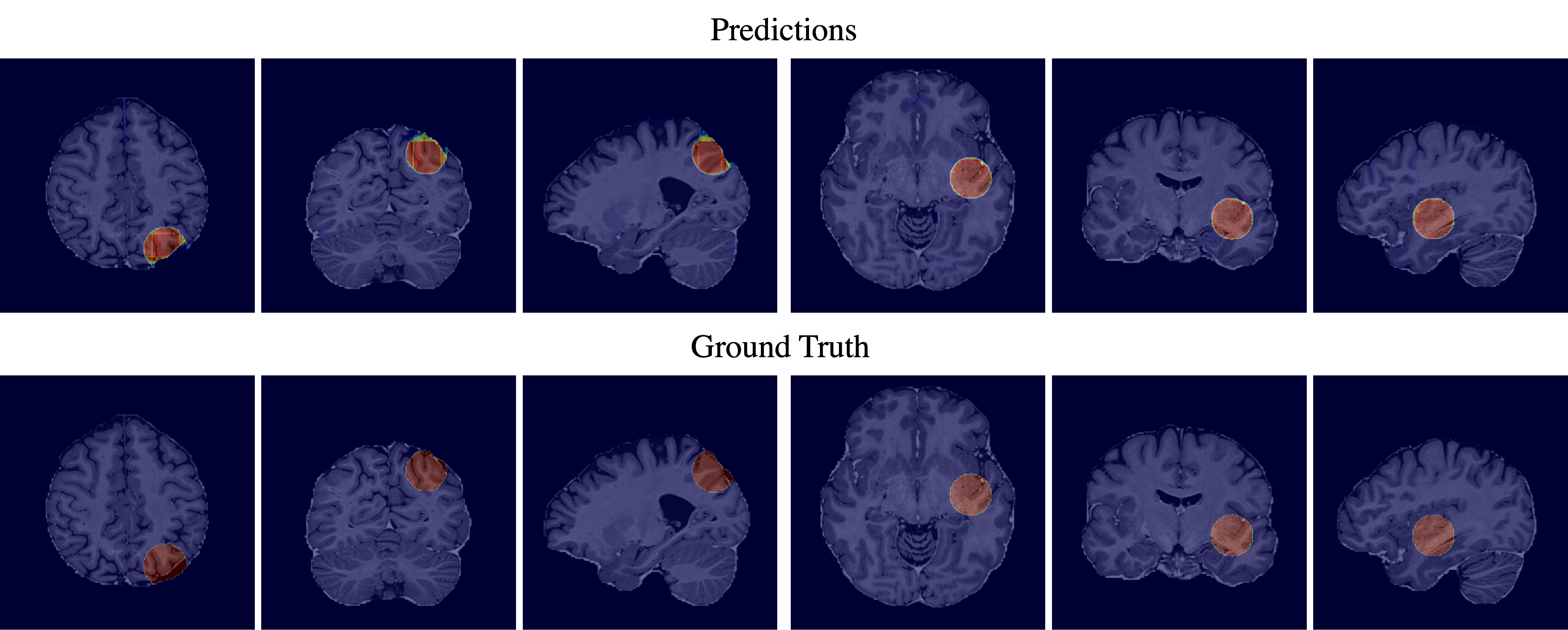}
    \caption{Predictions of our final model with AutoSeg and 2.5D training for artificial anomalies on the held-out test set of MOOD brain MRI. Left: uniform shift, right: noise addition.}
    \label{fig:mood_predictions}
\end{figure}

Figure \ref{fig:mood_predictions} displays the predicted anomaly maps and the ground truth segmentation maps of two random samples showing a \say{uniform shift} and a \say{noise addition} anomaly.
The predictions are almost perfect, showing only some minor inaccuracies at the borders.
Another slight error is visible in the random shift prediction. Here, the shift created a dent in the outer brain surface. The model predicts the anomaly only on the brain matter, missing the dent.

Figure \ref{fig:cxr14_cardiomegaly} shows the localization quality of our model for real-world anomalies on the example of cardiomegaly. Although being by far not as accurate as for artificial anomalies, the model is able to identify the heart as the source of the anomaly.

\section{Discussion}

Our proposed method steers the inductive bias of a model towards detecting pathologies -- instead of general statistical anomalies -- by generating artificial anomalies that better represent anomalies in the real world.
We have shown that our method produces impressive results on unseen artificial anomalies, outperforming existing methods by a large margin.
However, real-world pathologies have potentially very different properties.
This can be seen in the ChestX-ray14 dataset, where -- despite our model outperforming all competing methods -- the results are far from being usable, even in such a benchmark task. When faced with well-known practical problems like distribution shift, the performance might easily degenerate completely.
Likewise, the results in the MOOD challenge don't reflect the performance our model achieves on our artificial anomalies, indicating that samples in the hidden test set have different properties than our train anomalies.

Nevertheless, our results show that steering the inductive bias towards the actual goal of detecting pathologies improves their detection performance.
These findings motivate us to search for better anomalies with characteristics closer to real-world pathologies in future work.
%
%
%
%
\bibliographystyle{splncs04}
\bibliography{references}
\end{document}